\documentclass[12pt]{article}
\usepackage{stmaryrd}
\usepackage{bbm}
\usepackage{mathrsfs}
\usepackage{amsfonts}
\usepackage{amssymb}
\usepackage{amsmath}
\usepackage{amsthm}

\newtheorem{thm}{Theorem}[section]
\newtheorem{defi}{Definition}[section]

\def \part {\partial}
\def \be {\begin{equation}}
\def \ee {\end{equation}}
\def \bea {\begin{eqnarray}}
\def \eea {\end{eqnarray}}
\def \ba {\begin{array}}
\def \ea {\end{array}}

\def \R {\mathbb{R}}
\def \C {\mathbb{C}}

\numberwithin{equation}{section}

\allowdisplaybreaks[4]

\begin{document}

\title{ {\bf A refined invariant subspace method and applications to evolution equations}
\footnotetext{*Email: mawx@cas.usf.edu}}
\author{Wen-Xiu Ma\\
{\small Department of Mathematics and Statistics, University of
South Florida, }\\{\small Tampa, FL 33620-5700, USA} }
\date{}
\maketitle

{\bf Abstract:} The invariant subspace method is refined to present
more unity and more diversity of exact solutions to evolution
equations. The key idea is to take subspaces of solutions to linear
ordinary differential equations as invariant subspaces that
evolution equations admit. A two-component nonlinear system of
dissipative equations was analyzed to shed light on the resulting
theory, and two concrete examples are given to find invariant
subspaces associated with 2nd-order and 3rd-order linear ordinary
differential equations and their corresponding exact solutions with
generalized separated variables.

\vspace {5mm}



% {\bf 2000 MSC:} 37J15, 37K35, 35Q53, 35K55

\vspace {3mm}

\section{Introduction}

Exact solutions to differential equations are significantly
important in exploring the nature of motion. Based on the
classification of elementary functions, there are only three kinds
of explicit elementary exact solutions, classified as soliton-like,
positon-like and complexiton-like solutions
\cite{MaY-TAMS2005}-\cite{MaHL-NA2009}, to differential equations,
both linear and nonlinear. Hirota direct method presents Wronskian
and Pfaffian formulations of solutions to soliton equations, leading
to solitons, positons and complexitons
\cite{Hirota-book2004,Ma-DCDSS2009,MaAA-AMC2011}. For general
nonlinear partial differential equations, symmetry related methods
(see, e.g., \cite{FL}-\cite{QJW}) provide powerful approaches to
their exact solutions. Group-invariant solutions stemming from
symmetries play crucial roles in studying asymptotical behavior,
blow up phenomena and fractal properties of motion, and they can be
used to justify numerical schemes of solving partial differential
equations \cite{Blu,Olv,Debnath-book2004}.

The invariant subspace method, recently proposed in
\cite{Titov-book1988,G1}, is one of powerful approaches for
constructing exact solutions to nonlinear evolution equations.
Various invariant subspaces defined through linear ordinary
differential equations have been presented for solving specific
nonlinear evolution equations (see \cite{GS}-\cite{S1} and
references therein). Indeed, the invariant subspace method generates
many interesting exact solutions to nonlinear evolution equations in
mechanics and physics and a systematical solution procedure was
given by Galaktionov and Svirshchevskii in their book \cite{GS}.

In particular, Galaktionov \cite{G1} utilized the invariant subspace
method to generate exact solutions to nonlinear evolution equations
with quadratic nonlinearities, and showed that exact positive
solutions to the quasi-linear heat equations
$$u_t=(u^{-\frac43}u_x)_x-a u^{-\frac13}+b u^{\frac73}+c u
,$$ where $ a,b,c\in\mathbb{R}$ are constants, can be constructed
through invariant subspaces of functions of the polynomial or
trigonometric form, admitted by the spatial differential operator.
Actually, evolution equations that admit invariant subspaces can be
defined to be symmetries of given ordinary differential equations
\cite{S1,Svirshchevskii-CNSNS2004}. Interestingly, the $N$-soliton
solutions to soliton equations such as the KdV equation, the mKdV
equation, the nonlinear Schr\"{o}dinger equation and the sine-Gordon
equation, derived by the Hirota bilinear method
\cite{Hirota-book2004,HGR}, are all in a linear space of exponential
functions under change of variables \cite{GS}.

The invariant subspace method was also used to construct exact
solutions to systems of nonlinear evolution equations. On the basis
of the existence of invariant subspaces that systems of linear
ordinary differential equations define, Qu and Zhu \cite{QZ}
classified the systems of nonlinear parabolic equations of the form
\begin{align*}
& u_t=\left[f(u,v)u_x+p(u,v)v_x\right]_x+r(u,v),\\
&v_t=\left[g(u,v)u_x+q(u,v)v_x\right]_x+s(u,v).
\end{align*}
  Zhu and Qu \cite{ZQ} presented an estimation of maximal
  dimensions of invariant subspaces for two-component systems
  of nonlinear evolution equations, and Shen, Qu, Jin and Ji
\cite{SheQJJ-CAMB2011} generalized this estimation to
multi-component systems of nonlinear evolution equations, together
with some classifications of the considered systems of nonlinear
parabolic equations and computation of the exact solutions derived
from the corresponding invariant subspaces.

In this paper, we would like to refine the invariant subspace method
by taking invariant subspaces as subspaces of solution spaces to
systems of linear ordinary differential equations. Note that a
solution to an $n$th-order ordinary differential equation may not
satisfy another ordinary differential equation of order less than
$n$. Our idea will generalize the invariant subspace method from the
point of view of unity and diversity of invariant subspaces and
exact solutions. A two-component nonlinear system of dissipative
equations is analyzed carefully and a set of sufficient and
necessary conditions is presented for the existence of invariant
subspaces. Two concrete examples illustrate the effectiveness of the
resulting refined theory in presenting exact solutions with
generalized separated variables.

\section{Refining the invariant subspace method}

\subsection{Scalar case}

Let us consider a scalar evolution equation \be u_t=F[u],
\label{eq:ee:pma-rism}\ee
 where $u=u(x,t)$ is a function of $x,t\in \R $ and $F$ is a differential operator of order $m$:
 \be F[u] =F(x,t,u,u_1,\cdots , u_m),\ u_i=\frac {\part ^i} {\part  x^i}u,\ i\ge 0.\ee

Let $n\ge 1$ be a given natural number. Take $n$ linearly
independent functions \[f_1(x),f_2(x),\cdots , f_n(x),\] and form an
$n$-dimensional linear space \be W_n={\mathcal L}\{f_1(x)
,f_2(x),\cdots ,f_n(x)\}=\bigl\{\sum_{i=1}^n C_if_i(x)\, |\,
C_i=\textrm{const.},\ 1\le i\le n \bigr\}, \ee i.e., the linear span
of $f_1(x),f_2(x),\cdots , f_n(x)$ over $\R $ or $\C$.

\begin{defi}
A finite-dimensional linear space $ W_n$ is said to be invariant
with respect to a differential operator $F$, if $F[W_n]\subseteq
W_n$, i.e., $F[u]\in W_n,\ \forall u\in W_n$.
\end{defi}

Suppose that $W_n$ is invariant with respect to a given differential
operator $F$. Then there exist $n$ functions $\tilde F_1$, $\tilde
F_2$, $\cdots$, $\tilde F_n$ such that \be F\bigl[\sum_{i=1}^n
C_if_i(x)\bigr]=\sum_{i=1}^n \tilde F_i(C_1,C_2,\cdots, C_n)f_i(x)
\ee for whatever constants $C_1,C_2,\cdots, C_n $. It follows that
the evolution equation \eqref{eq:ee:pma-rism} possesses a solution
of the form \be u(x,t)=\sum_{i=1}^n \phi_i(t)f_i(x),
\label{eq:solutionform:pma-rism} \ee
 if and only if  $\phi_1,\phi_2,\cdots,\phi_n$ satisfy
 a system of ordinary differential equations:
\be  \frac {d\phi _i}{dt }=\tilde F_i(\phi_1,\phi_2,\cdots, \phi_n),
\ 1\le i\le n. \ee

We usually take an invariant subspace $W_n$ as the space of
solutions to a given $n$th-order linear ordinary differential
equation: \be L[y]=y^{(n)}+a_{n-1}(x)y^{(n-1)}+\cdots +a_0(x)y=0,\
y^{(i)}=D^iy,\ D=\frac d {dx} ,\ i\ge 0, \label{eq:lode:pma-rism}\ee
 where $a_0,a_1,\cdots ,a_{n-1}$ are given continuous functions. The linearity of the above equation brings a good
 possibility to generate exact solutions to nonlinear evolution equations.

The above approach for constructing exact solutions of the form
\eqref{eq:solutionform:pma-rism} is called the invariant subspace
method \cite{Titov-book1988,G1}. It is also called a generalized
separation of variables \cite{S1}, whose resulting solutions of the
form \eqref{eq:solutionform:pma-rism} we call solutions with
generalized separated variables.

To refine the invariant subspace method discussed above, let us
consider a $k$-dimensional subspace $W_k$ of the $n$-dimensional
linear space $W_n$, and without loss of generality, we set \be
W_k=\mathcal{L}\{f_1(x) ,f_2(x),\cdots ,f_k(x)\}=\bigl\{\sum_{i=1}^k
C_if_i(x)\, | \,C_i=\textrm{const.},\ 1\le i\le k \bigr\}, \ee where
$k\le n$. The invariance condition $F[W_k]\subseteq W_k$ means that
there exist $k$ functions $\bar F_1,\bar F_2,\cdots ,\bar F_k$ such
that \be F\bigl[\sum_{i=1}^k C_if_i(x)\bigr]=\sum_{i=1}^k \bar
F_i(C_1,C_2,\cdots, C_k)f_i(x) \ee for whatever constants
$C_1,C_2,\cdots, C_k $. This way, beginning with a similar system of
ordinary differential equations \be  \frac {d\psi _i}{dt }=\bar
F_i(\psi_1,\psi_2,\cdots, \psi_k), \ 1\le i\le k, \ee we can
engender a set of exact solutions to the evolution equation
\eqref{eq:ee:pma-rism}: \be u= \sum_{i=1}^k\psi _i(t) f_i(x).\ee We
call this approach a refined invariant subspace method. One of its
advantages is that when $k<n$, the invariance condition
$F[W_k]\subseteq W_k$ requires much less conditions on the evolution
equation \eqref{eq:ee:pma-rism}.

The following results are helpful in searching for conditions to
guarantee the existence of invariant subspaces that nonlinear
evolution equations admit.

\begin{thm}\label{thm:linearindependenceoffunctions:pma-rism}
Let $I=(a,b) $ be an open interval, $x_0\in I$ be an arbitrary point
and $m\ge 0$ be an integer. If a real function $f$ defined on $I$
satisfies \be f(x_0)=0,\ f'(x_0)=0,\ \cdots, \ f^{(m-1)}(x_0)=0,\
f^{(m)}(x_0)=d, \label{eq:conditionforderivatives:pma-rism}\ee where
$d $ is nonzero, then
 $ f,f',\cdots ,f^{(m)}$ are
linearly independent over $I$.
\end{thm}

\indent {\it Proof}: If $m=0$, the theorem is true, since based on
\eqref{eq:conditionforderivatives:pma-rism}, $f$ is not a zero
function. In what follows, let $m>0$. Suppose that the theorem is
not true. That is to say, there are an integer $0\le k\le m$ and
real constants $c _i,\ 0\le i\le m-k$, with $c_{m-k}\ne 0$ such that
\be c_0f(x)+c_1f'(x)+\cdots + c_{m-k}f^{(m-k)}(x)=0,\ x\in
I.\label{eq:lineardependenceforfunctions:pma-rism} \ee
  This is an ordinary differential equation with constant coefficients. Thus, the solution $f$ is analytic in $I$, and so, using the conditions for derivatives in \eqref{eq:conditionforderivatives:pma-rism}, we have
  \be f(x)= \frac d {m!} (x-x_0)^{m}+\sum_{i=m+1}^\infty \frac {f^{(i)}(x_0)}{i!}(x-x_0)^i.\ee
Now, the coefficient of $(x-x_0)^{k}$ in the Taylor series expansion
of the function on the left-hand side of
\eqref{eq:lineardependenceforfunctions:pma-rism}
 is
$ {c_{m-k} d}/{k!},$ which is not zero. This contradicts the linear
dependence equation
\eqref{eq:lineardependenceforfunctions:pma-rism}. Therefore, $ f, f'
, \cdots , f^{(m)} $ are linearly independent over $I$. \hfill $\Box
$

\begin{thm}\label{thm:linearindependence:pma-rism}
Let $a_i(x),\ 0\le i\le n-1,$ be real continuous functions on an
open interval $I=(a,b)$. Then there exits a solution $y$ to the
linear ordinary differential equation
\[ y^{(n)}+a_{n-1}(x) y^{(n-1)} +\cdots + a_0(x) y=0  \]
 such that
 $ y,y',\cdots ,y^{(n-1)}$ are
linearly independent over $I$.
\end{thm}

\indent {\it Proof}: For a fixed $x_0\in I$, let us consider a
Cauchy problem of the linear ordinary differential equation
\eqref{eq:lode:pma-rism} with the initial data: \be y(x_0)=0,\
y'(x_0)=0,\ \cdots, \ y^{(n-2)}(x_0)=0,\ y^{(n-1)}(x_0)=d,
\label{eq:initialdataforode:pma-rism}\ee where the real constant $d
$ is nonzero. The theory of ordinary differential equations tells
that there exits a unique solution $y$ defined on $I$ to this Cauchy
problem.
 It now follows from Theorem \ref{thm:linearindependenceoffunctions:pma-rism} with $m=n-1$ that this solution $y$ is the desired solution.
\hfill $\Box $

 Theorem \ref{thm:linearindependence:pma-rism}
tells that for an $n$th-order linear ordinary differential equation,
there always exists a solution $y$ such that $y,y',\cdots,
y^{(n-1)}$ are linearly independent. Thus, in principle, for a given
differential operator $F$, we can get sufficient and necessary
conditions to guarantee the existence of an invariant subspace
$W_k$, by collecting all coefficients of linearly independent terms,
generated from the linearly independent functions $y,y',\cdots
,y^{(n-1)}$, in the invariance condition \be
D^nF[y]+a_{n-1}(x)D^{n-1}F[y]+\cdots + a_0(x) F[y]=0,\ y\in W_k,
\label{eq:invariancecondition:pma-rism} \ee and setting them to be
zero. The process may be difficult and frustrating. However, from
analyzing different terms involving $y,y',\cdots, y^{(n-1)}$ in
 \eqref{eq:invariancecondition:pma-rism},
we can always get sufficient conditions for the existence of
invariant subspaces that $F$ admits.

It should be interesting to note that $W_k$ may not possibly be
generated by a $k$th-order linear ordinary differential equation. An
example is given as follows:
\[W_1={\mathcal L}\{ y \},\ y=e^{\lambda _1x}+e^{\lambda_2 x},\ \lambda_{1,2}=\textrm{consts.},\] where $y$ is a solution to the 2nd-order linear differential equation
\[ y''-(\lambda _1 +\lambda _2)y'+\lambda _1\lambda _2y=0, \]
but doesn't solve any 1st-order linear differential equation when
$\lambda _1\ne \lambda _2$. Therefore, our refinement does make
sense in generalizing the invariant subspace method.

\subsection{Multi-component case}

In what follows, we adopt the notations
\begin{equation}
u_0^i=u^i(x,t),\ u_j^i=\frac{\partial^ju^i(x,t)}{\partial x^j}, \
1\le i\le q, \  j\ge 1, \end{equation} such that the discussion can
be easily extended to cases of multiple spatial variables. A system
of evolution equations is assumed to take the form
\begin{eqnarray}\label{eq:mee:pma-rism}
u_t=F[u]= (F^1[u], F^2[u], \cdots, F^q[u])^T,\ u=(u^1,u^2,\cdots, u^q)^T,
\end{eqnarray}
 where
\begin{eqnarray}\label{eq:degreesofmee:pma-rism}
F^i[u]=F^i(x,t, u^1, \cdots,u^q, \cdots, u^1_{m_i},\cdots,  u^q_{m_i} ), \ 1\le i\le q,
\end{eqnarray}
are given sufficiently smooth functions in the indicated variables.
Therefore, for each $1\le i\le q$, $F^i$ can be viewed as a
differential operator of order $m_i$.

Let ${W}_{k_1,\cdots,k_q}$ denote a linear space $W^1_{k_1}\times
\cdots \times W^q_{k_q} $, with $W^i_{k_i}$ being defined by
\begin{eqnarray}
W_{k_i}^i=\mathcal{L} \{f_1^i(x), \cdots, f_{k_i}^i(x)\}=
\bigl\{\sum_{j=1}^{k_i}C_j^if^i_j(x) \,|\, C_j^i=\textrm{const.},\ 1\le j\le k_i \bigr\}, \ 1\le i\le q,
\end{eqnarray}
where for each $1\le i\le q$, $f_1^i(x), \cdots, f_{k_i}^i(x)$ are
linearly independent. If the above vector differential operator
${F}$ satisfies the invariance condition
\begin{eqnarray*}
{F}[u]\in W_{k_1,\cdots,k_q},\ \forall u\in W_{k_1,\cdots,k_q},
\end{eqnarray*}
namely,
\begin{eqnarray}
F^i[u]\in W^i_{k_i} ,\ \forall u\in  W_{k_1,\cdots,k_q},\ 1\le i\le q, \label{eq:generalinvariancecond:pma-rism}
\end{eqnarray}
then the vector differential operator ${F}$ (or the system of
evolution equations \eqref{eq:mee:pma-rism}) is said to admit an
invariant subspace ${W}_{k_1,\cdots,k_q}$, or ${W}_{k_1,\cdots,k_q}$
is said to be invariant under the given differential operator ${F}$.
The above invariance condition
\eqref{eq:generalinvariancecond:pma-rism} means that there exist
functions $\tilde F^i_j$, $1\le j\le k_i,\ 1\le i \le q$, such that
\begin{eqnarray}
F^i\bigl[\sum_{j=1}^{k_1}C_j^1f_j^1(x), \cdots,
\sum_{j=1}^{k_q}C_j^qf_j^q(x)
\bigr]=\sum_{j=1}^{k_i}\tilde F_j^i(C_1^1,\cdots,C_{k_1}^1, \cdots,
C_1^q, \cdots, C_{k_q}^q)f_j^i(x),
\end{eqnarray}
where $1\le i\le q$.

Now if a space ${W}_{k_1,\cdots,k_q}$ is admitted by the vector
differential operator ${F}$, then the system of evolution equations
\eqref{eq:mee:pma-rism} possesses an exact solution of the form
\begin{eqnarray}
u^i=\sum_{j=1}^{k_i}C_j^i(t) f_j^i(x), \  1\le i\le q,
\end{eqnarray}
if and only if the $C_j^i(t)$'s satisfy a system of ordinary
differential equations:
\begin{eqnarray}
\frac{{ d}C_j^i}{{ d}t}=\tilde F^i_j (C_1^1, \cdots,
C_{k_1}^1,\cdots C_1^q, \cdots, C_{k_q}^q ),\
1\le j\le k_i,\  1\le i\le q.
\end{eqnarray}

The last step is that for each $1\le i\le q$, we take the space
$W_{k_i}^i=\mathcal{L}\{f_1^i(x), \cdots, f_{k_i}^i(x) \}$ as a
subspace of solutions to an $n_i$th-order linear ordinary
differential equation:
\begin{eqnarray}\label{eq:mlinearODEs:pma-rism}
L_i[y_i]=y_i^{(n_i)}+a_{n_i-1}^i(x)y_i^{(n_i-1)}+\cdots+a_1^i(x)y_i'+a_0^i(x)y_i=0,
\end{eqnarray}
where $ n_i\ge k_i$. The invariance conditions for the subspace
${W}_{k_1,\cdots,k_q}=W_{k_1}^1\times \cdots \times W_{k_q}^q$ with
respect to ${F}=(F^1,\cdots,F^q)^T$ read
\begin{eqnarray}\label{eq:minvarianceconditions:pma-rism}
D^{n_i} F^i[u]+a_{n_i-1}^i(x) D^{n_i-1}F^i[u]+\cdots +a_0^i(x)F^i[u]=0,\ u\in W_{k_1,\cdots,k_q},\ 1\le i\le q.
\end{eqnarray}
This set of equations is our starting point to construct exact
solutions to systems of evolution equations by looking for their
invariant subspaces.

Note that the orders of linear ordinary differential equations
defining invariant subspaces can not be arbitrary, and they are
subject to the differential orders of the nonlinear operators $F^i,\
1\le i\le q$. Once the maximal orders of the required linear
ordinary differential equations are determined, we will be able to
classify systems of evolution equations under consideration, and
compute exact solutions from the associated invariant subspaces.

The problem of maximal orders of linear ordinary differential
equations defining invariant subspaces was firstly posed and solved
for the scalar case in \cite{GS}. For the scalar case, the maximal
order of a linear ordinary differential equation defining an
invariant subspace is not greater than $2m+1$, where $m$ is the
order of the differential operator $F$ in \eqref{eq:ee:pma-rism}.
For a $q$-component nonlinear really-coupled system defined by
\eqref{eq:mee:pma-rism} and \eqref{eq:degreesofmee:pma-rism} with
$m_1\ge m_2\ge \cdots \ge m_q\ge 0$, the orders $\{n_1,\cdots,n_q\}$
of linear ordinary differential equations defining invariant
subspaces with $n_1\ge  n_2\ge \cdots \ge n_q>0$ must satisfy
\cite{SheQJJ-CAMB2011}:
\begin{eqnarray*} n_{i-1}-n_i\leq m_i,
\  2\le i \le q, \  n_1\le  2 \sum
_{i=1}^q m_i +1.
\end{eqnarray*}
That the system defined by \eqref{eq:mee:pma-rism} and
\eqref{eq:degreesofmee:pma-rism} is really-coupled means that for
each pair $1\le i\ne j\le q$, there exists an integer $0\le k\le
m_i$ such that $\frac {\partial F^i}{\partial u^j_{k}}$ is not a
zero function. If $F^i,\ 1\le i\le q$, are all real, then the
really-coupled condition can be concisely written as
\[  \sum_{k=0} ^{m_i}\bigl(\frac {\partial F^i}{\partial u^j_{k}}\bigr)^2\ne 0, \  1\le i\ne j\le q.\]

\section{Invariant subspaces and exact solutions}

In this section, we analyze a (1+1)-dimensional nonlinear system of
dissipative equations
 to illustrate how
to generate invariant subspaces and the corresponding exact
solutions. We consider the following nonlinear system of dissipative
equations:
\begin{eqnarray}
&& u_t= F =(u_{xx}+ \alpha _1  vv_x)_x+ \alpha _2 v^2,\\
&& v_t= G =u_{xx}+ \beta _1  u+ \beta _2 v,
\end{eqnarray}
where $\alpha _1,\alpha _2,\beta _1,\beta_2$ are constants, $ \alpha
_1, \alpha _2 $ are not simultaneously equal to zero, and we have
used the traditional notation
\[u_x=\frac {\partial u}{\partial x},\ v_x=\frac {\partial v}{\partial x},\ u_{xx}=\frac {\partial ^2 u}{\partial x^2},\ v_{xx}=\frac {\partial ^2 v}{\partial x^2},\  \cdots.\]

Let us take an invariant subspace $W_{2,2}=W_2^1\times W_2^2$
defined by \be L_1[y]=y''+a_1y'+a_0y=0,\ L_2[z]=z''+b_1z'+b_0z=0,\ee
where $a_0,a_1,b_0,b_1$ are constants to be determined. The
corresponding invariance conditions read
\begin{eqnarray}
&&(D^2F +a_1D F  +a_0 F )\big|_{u\in W^1_2,\,v\in W_2^2}=0,\label{eq:eq1ofexmple1ofinvariancecondition:pma-rism}\\
&&(D^2 G +b_1D G +b_0 G )\big|_{u\in W^1_2,\,v\in W_2^2}=0.\label{eq:eq2ofexmple1ofinvariancecondition:pma-rism}
\end{eqnarray}

Substitute the expressions for $ F $ and $ G$ into the above
equations, and replace $u_{xx}$ and $v_{xx}$ by $-a_1u_x-a_0u $ and
$-b_1v_x-b_0v$ a few times, respectively. Then we collect the
coefficients of $(v_x)^2,vv_x$ and $v^2$ in the first simplified
equation and the coefficients of $u_x$ and $u$ in the second
simplified equation, and set them to be zero, to obtain the
sufficient conditions:
\begin{align}
(v_x)^2:\quad &7 \alpha _1 {b_1}^2  + \alpha _1 a_0   +2 \alpha _2  -4 \alpha _1 b_0  -3 \alpha_1 a_1  b_1  =0,\label{eq:cond1:pma-rism}\\
vv_x:\quad &12 \alpha _1 b_1b_0- \alpha _1 a_0b_1-4\alpha _1 a_1 b_0+2\alpha _2 a_1-\alpha _1 {b_1}^3-2\alpha _2 b_1+\alpha _1 a_1{b_1}^2=0,\label{eq:cond2:pma-rism}\\
v^2:\quad &4\alpha _1 {b_0}^2+\alpha _2 a_0-\alpha _1 {b_1}^2b_0+\alpha _1 a_1b_0 b_1-\alpha _1 a_0b_0-2\alpha _2 b_0 =0,\label{eq:cond3:pma-rism}\\
u_x:\quad &-{a_{{1}}}^{3}+2\,a_{{0}} a_{{1}}-\beta _1 a_{{1}}+{a_{{1}}}^{2}b_{{1}} -a_{{0}} b_{{1}}
+\beta _1 b_{{1}}-a_{{1}} b_{{0}}
=0, \label{eq:cond4:pma-rism}\\
u:\quad & -a_{{0}} {a_{{1}}}^{2}+{a_{{0}}}^{2}-\beta _1 a_{{0}}+a_{{0}}a_{{1}} b_{{1}}-a_{
{0}}b_{{0}}+\beta _1 b_{{0}}
 =0,\label{eq:cond5:pma-rism}
 \end{align}
which guarantees the invariance conditions
\eqref{eq:eq1ofexmple1ofinvariancecondition:pma-rism} and
\eqref{eq:eq2ofexmple1ofinvariancecondition:pma-rism}. We began with
two second order differential equations, and so definitely there
exist linearly dependent terms in $(v_x)^2,vv_x$ and $v^2$ for
whatever solution $v$, but $u$ and $u_x$ could be linearly
independent (see Theorem \ref{thm:linearindependence:pma-rism}).
Therefore, the conditions
\eqref{eq:cond1:pma-rism}-\eqref{eq:cond3:pma-rism} are sufficient
but not necessary to guarantee the first invariance condition
\eqref{eq:eq1ofexmple1ofinvariancecondition:pma-rism}, but the
conditions \eqref{eq:cond4:pma-rism} and \eqref{eq:cond5:pma-rism}
are both sufficient and necessary to guarantee the second invariance
condition \eqref{eq:eq2ofexmple1ofinvariancecondition:pma-rism}.

Under the first condition \eqref{eq:cond1:pma-rism}, the second and
third conditions, \eqref{eq:cond2:pma-rism} and
\eqref{eq:cond3:pma-rism}, are equivalent to \bea && -  \alpha _1
a_{{1}} {b_{{1}}}^{3}-2\, \alpha _1 b_{{0}} {b_{{1}}}^{2}+3\,\alpha
_1 {b_{{1}}}^{4}-\alpha _2 a_{{0 }}+  \alpha _2 a_{{1}} b_{{1}}
 =0,\label{eq:simplecond1:pma-rism}\\
&& -3\,\alpha _1 {b_{{1}}}^{3}-\alpha _2 a_{{1}}+ \alpha _1 a_{{1}}
{b_{{1}}}^{2}+2\, \alpha _1 a_{{1}} b_{{0}}-4 \,\alpha _1
b_{{0}}b_{{1}} =0.\label{eq:simplecond2:pma-rism}
 \eea
This can be shown directly by using Maple and we will see later
where they come from.

Let us now assume $\Delta _2={b_1}^2-4b_0>0$. Then \be  W_2^2
={\cal L}\, \{ e^{\lambda _+ x},\,e^{\lambda _{-}x} \},\ee where \[
\lambda _\pm =\frac {-b_1\pm \sqrt{\Delta _2}}2.\] Collecting the
coefficients of three linearly independent terms $e^{(\lambda
_++\lambda _{-})x}$, $e^{2\lambda _+ x}$ and $e^{2\lambda _{-} x}$
in the first invariance condition
\eqref{eq:eq1ofexmple1ofinvariancecondition:pma-rism} and setting
them to be zero
 gives rise to
\be \gamma_1 =0,\ \gamma _2 \pm \sqrt{\Delta _2} \, \gamma _3
=0,\label{eq:sufficientandnecessaryinvariancecondition:pma-rism} \ee
respectively, where \bea && \gamma_1=\alpha _1 a_{{0}}{b_{{1}}}^{2}-
\alpha _1 a_{{1}}{b_{{1}}}^{3}-2\,\alpha _2 a_{{1}} b_{{1}}+\alpha
_1 {b_{{1 }}}^{4}+2\,\alpha _2 {b_{{1}}}^{2}+2\,\alpha _2 a_{{0}} ,
\\
&& \gamma_2=-4\,\alpha _2 b_{{0}}+\alpha _2 a_{{0}}+8\,\alpha _1
{b_{{0}}}^{2}-16\,\alpha _1 b_0 {b_{{1}}}^{2}-2\,\alpha -1 a
_{{0}} b_{{0}}+4\,\alpha _1 {b_{{1}}}^{4}\nonumber \\
&& \qquad +2\,\alpha _2 {b_{{1}}}^{2}+6\,\alpha _1 a_{{1}}b_{{0}}b
_{{1}}-2\,\alpha _1 a_{{1}}{b_{{1}}}^{3}-\alpha _2 a_{{1}}
b_{{1}}+\alpha _1 a_{{0}}{b_{{1}}}^{2} ,
\\
&& \gamma_3=-4\,\alpha _1 {b_{{1}}}^{3}-2\,\alpha _2 b_{{1}}+\alpha
_2 a_{{1}}+2\,\alpha _1 a_{{1}}{b_{{1}}}^{2}-2\,\alpha _1 a_
{{1}}b_{{0}}-\alpha _1 a_{{0}}b_{{1}}+8\,\alpha _1 b_{{0}}b_{{1}} .
 \eea
The conditions in
\eqref{eq:sufficientandnecessaryinvariancecondition:pma-rism}
 equivalently lead to
 \be
 \gamma_1=0,\ \gamma_2=0,\ \gamma _3=0.
 \label{eq:simplilfiedsufficientandnecessaryinvariancecondition:pma-rism}\ee
These are sufficient and necessary conditions for guaranteeing the
first invariance condition
\eqref{eq:eq1ofexmple1ofinvariancecondition:pma-rism}.

Under the first condition \eqref{eq:cond1:pma-rism}, $\gamma_2=0$
and $\gamma _3=0$ become the equations
\eqref{eq:simplecond1:pma-rism} and \eqref{eq:simplecond2:pma-rism},
respectively. When the three equations, \eqref{eq:cond1:pma-rism},
$\gamma _2=0$ and $\gamma _3=0$, hold, the condition $\gamma _1=0$
is automatically satisfied. Therefore, the conditions
\eqref{eq:cond1:pma-rism}, \eqref{eq:cond2:pma-rism} and
\eqref{eq:cond3:pma-rism} yield
\eqref{eq:simplilfiedsufficientandnecessaryinvariancecondition:pma-rism}.
But conversely, it is not true. This is because the system
\eqref{eq:simplilfiedsufficientandnecessaryinvariancecondition:pma-rism}
has a solution $\alpha _2= b_0=b_1=0$
 with the other variables being arbitrary, but the left-hand side of \eqref{eq:cond1:pma-rism} under $\alpha _2= b_0=b_1=0$ is $ \alpha _1 a_0$,
 not always zero.

More specifically, under $\Delta _2={b_1}^2-4b_0>0$, let us take a
smaller invariant subspace: \be W_{2,1}=W_2^1\times W_{1}^2,\
W_1^2={\cal L} \{e^{\mu x}\}, \ee where $\mu=\lambda _+ $ (or
$\mu=\lambda _{-} $). Then the invariance conditions
\begin{eqnarray}
&&(D^2F +a_1D F  +a_0 F )\big|_{u\in W^1_2,\,v\in W_1^2}=0,\label{eq:eq1ofexmple2ofinvariancecondition:pma-rism} \\
&&(D^2 G +b_1D G +b_0 G )\big|_{u\in W^1_2,\,v\in W_1^2}=0,\label{eq:eq2ofexmple2ofinvariancecondition:pma-rism}
\end{eqnarray}
only require \be \gamma _2+ \sqrt{\Delta _2}\, \gamma _3=0 \ ({\rm
or}\ \gamma _2 - \sqrt{\Delta _2}\, \gamma
_3=0)\label{eq:example2ofinvariancecondition:pma-rism}\ee plus
\eqref{eq:cond4:pma-rism} and \eqref{eq:cond5:pma-rism}. Any of
these two conditions in
\eqref{eq:example2ofinvariancecondition:pma-rism} is much weaker
than the conditions in
\eqref{eq:sufficientandnecessaryinvariancecondition:pma-rism}, i.e.,
\eqref{eq:simplilfiedsufficientandnecessaryinvariancecondition:pma-rism}.
Therefore, we can have a more general system of dissipative
equations which still possesses exact solutions with generalized
separated variables.

In what follows, we give two concrete examples of getting exact
solutions with generalized separated variables.

\vskip 2mm

\noindent {\bf Example 1:} Let us consider a system
\begin{eqnarray}
&&u_t=(u_{xx}+ \alpha _1  vv_x)_x+(3\alpha _1 a_1  b_1-\frac{9}{2} \alpha _1 {b_1}^2-\frac 12 \alpha_1 {a_1}^2) v^2, \\
&&v_t=u_{xx} + \beta _1 u+ \beta_2 v,
\end{eqnarray}
which admits an invariant subspace $W_{2,2}$ defined by
\bea &&L_1[y]= y''+a_1 y'+(a_1b_1-{b_1}^2)y=0, \\
&& L_2[z]=z''+b_1 z'+(a_1b_1-\frac 34 {b_1}^2-\frac 14 {a_1}^2)z=0,
\eea where $a_1$ can take any of the following three choices: \be
a_1=b_1\ {\rm or }\ a_1={\frac {4\,{b_{{1}}}^{2}-\beta _1 \pm \sqrt
{4\,{b_{{1}}}^{4}-20\,\beta _1 {b_{{1}}}^ {2}+{\beta _1
}^{2}}}{6b_{{1}}}} .\ee

We analyze the case of $a_1=b_1$, for which we have
\begin{eqnarray}
&&u_t=(u_{xx}+ \alpha _1 vv_x)_x- 2\alpha _1 {a_1}^2  v^2, \label{eq:eq1ofconcreteexample1:pma-rism}
\\
& &
v_t=u_{xx} + \beta _1 u+ \beta _2 v,   \label{eq:eq2ofconcreteexample1:pma-rism}
\end{eqnarray}
and \bea &&
L_1[y]=y''+a_1 y'=0,\\
&& L_2[z]=z''+a_1 z'=0. \eea
 From these two equations $L_1[y]=0$ and $ L_2[z]=0$, we obtain an invariant
subspace
\begin{align}
W_2^1\times W_2^2=\mathcal{L}\left\{1, \,e^{-a_1 x}
\right\}\times
\mathcal{L}\left\{1,\,e^{-a_1 x}\right\}
\end{align}
that the system of \eqref{eq:eq1ofconcreteexample1:pma-rism} and
\eqref{eq:eq2ofconcreteexample1:pma-rism} admits. It then follows
that an exact solution takes the form
\begin{align}\label{eq:exactsolutionofconcreteexample1:pma-rism}
&u=C_1(t)+C_2(t)e^{-a_1x},\ v=D_1(t)+D_2(t)e^{-a_1 x}.
\end{align}
Substituting this solution into the system of
\eqref{eq:eq1ofconcreteexample1:pma-rism} and
\eqref{eq:eq2ofconcreteexample1:pma-rism}, we get the following
system of ordinary differential equations:
 \bea
&& C_1'=-2 \alpha _1  {a_1}^2{D_1}^2, \ C_2'= -{a_1}^3C_2 -3 \alpha
_1 {a_1}^2 D_1 D_2,\label{eq:Ceq1ofconcreteexample1:pma-rism}
\\
&& D_1'=\beta _1 C_1+\beta _2 D_1,\ D_2'={a_1}^2 C_2 +\beta _1  C_2
+ \beta _2 D_2.\label{eq:Ceq2ofconcreteexample1:pma-rism}
 \eea

Based on the refined theory, let us further focus on a smaller
invariant subspace \be W_{1,1}=W_1^1\times W_{1}^2={\mathcal
L}\{e^{-a_1x}\}\times {\mathcal L}\{e^{-a_1x}\},\ee to present exact
solutions. Solving the system of
\eqref{eq:Ceq1ofconcreteexample1:pma-rism} and
\eqref{eq:Ceq2ofconcreteexample1:pma-rism} with $C_1=D_1=0$, we
arrive at
 \be
C_2=c\,e^{-{a_1}^3t}, \ D_2=d \, e^{\beta _2 t}- \frac{c
({a_1}^2+\beta _1 )}{{a_1}^3+\beta _2 }\, e^{-{a_1}^3t}, \ee and
then according to
\eqref{eq:exactsolutionofconcreteexample1:pma-rism}, we obtain an
exact solution to the system of
\eqref{eq:eq1ofconcreteexample1:pma-rism} and
\eqref{eq:eq2ofconcreteexample1:pma-rism}: \be  u= c \, e^{-a_1 x
-{a_1}^3t},\ v= d\, e^{-a_1 x+\beta _2 t} -\frac {c ({a_1}^2+\beta
_1 )}{{a_1}^3+\beta _2 } \, e^{-a_1 x -{a_1}^3t},\ee where $c$ and
$d$ are arbitrary constants.

\vskip 2mm

\noindent {\bf Example 2:} Let us finally consider a system of the
following evolution equations:
\begin{eqnarray}
&&u_t=(u_{x}+ \alpha _1  vv_x)_x-{a_1}^2\alpha_1  v^2 -3 \alpha_1 vv_{xx}, \label{eq:eq1ofconcreteexample2:pma-rism}\\
&&v_t=u_{xx} + {a_1}^2 u+ \beta _2 v,\label{eq:eq2ofconcreteexample2:pma-rism}
\end{eqnarray}
where $a_1\ne 0,\alpha _1,\beta_2$ are arbitrary constants. This
system admits an invariant subspace $W_{1,1}$ defined by \be
W_{1,1}=W_1^1\times W_1^2={\mathcal L}\{\cos (a_1 x)\} \times
{\mathcal L}\{ 1+\sin (a_1 x)\}. \ee These two basis solutions
$y=\cos( a_1 x)$ and $z=1+\sin (a_1x)$ satisfy
\bea &&L_1[y]= y''+{a_1}^2 y=0, \\
&& L_2[z]=z'''+{a_1}^2 z'=0, \eea but they cannot satisfy any
lower-order linear ordinary differential equations with constant
coefficients.

Now, assuming a solution with the form \be u=C(t) \cos (a_1 x),\
v=D(t) [1+\sin (a_1x)], \ee and substituting back into the system of
\eqref{eq:eq1ofconcreteexample2:pma-rism} and
\eqref{eq:eq2ofconcreteexample2:pma-rism}, we find \be C'(t)=
-{a_1}^2C(t),\ D'(t)=\beta _2 D(t).\ee The general solution of this
system yields the following exact solution to the system of
\eqref{eq:eq1ofconcreteexample2:pma-rism} and
\eqref{eq:eq2ofconcreteexample2:pma-rism}: \be
u=c\,e^{-{a_1}^2t}\cos (a_1 x),\ v=d\, e^{\beta _2 t}[1+\sin
(a_1x)], \ee where $c$ and $d$ are two arbitrary constants.

This gives us a concrete example which uses the refined invariant
subspace method to construct exact solutions to nonlinear systems of
evolutions equations.

\section{Concluding remarks}

The invariant subspace method was refined by taking invariant
subspaces as subspaces of solutions to linear ordinary differential
equations. Our discussions were concentrated on how to identify
sufficient and necessary conditions for the existence of invariant
subspaces that nonlinear evolution equations admit. Two concrete
examples illustrated the effectiveness of the refined approach for
exploring solution structures of systems of nonlinear differential
equations, which are notoriously more difficult to solve than scalar
ones.

The invariant subspace method is also called a generalized
separation of variables for nonlinear differential equations by
Svirshchevskii \cite{S1}, presenting a kind of complexiton-like
solutions \cite{Ma-NA2005,Ma-PLA2002}. It is interesting to see that
the linear superposition principle takes on key role in constructing
exact solutions to either evolution equations \cite{GS} or Hirota
bilinear equations \cite{MaF-CMA2011}. All related theories furnish
linear combination solutions of functions with separated variables,
which shows a sort of integrability of nonlinear differential
equations \cite{Ma-book2005}.

Motivated by the multiple exp-function method \cite{MaHZ-PS2010}, we
can also characterize invariant subspaces through
 linear partial differential equations. This will lead to more diverse situations of solutions with generalized separated variables and open a much larger research area.

\vskip 2mm

 \small
 {\bf Acknowledgements:}  The work was supported in part by the State Administration of
Foreign Experts Affairs of China, the National Natural Science
Foundation of China (Nos. 10971136, 10831003, 61072147 and
11071159),
 Chunhui Plan of the Ministry of Education of China,
 Zhejiang Innovation Project (Grant No. T200905), the
Natural Science Foundation of Shanghai and the Shanghai Leading
Academic Discipline Project (No. J50101). The author is also very
grateful to Alrazi Abdeljabbar, Magdy G. Asaad,
 Boris Shekhtman, Junyi Tu and Yuan Zhou for helpful discussions.

\end{document}